# Boundary knot method for Laplace and biharmonic problems

**W. Chen**
Department of Informatics, University of Oslo, P.O.Box 1080, Blindern, 0316 Oslo, Norway
E-mail: wenc@ifi.uio.no

## ABSTRACT

**Summary** The boundary knot method (BKM) [1] is a meshless boundary-type radial basis function (RBF) collocation scheme, where the nonsingular general solution is used instead of fundamental solution to evaluate the homogeneous solution, while the dual reciprocity method (DRM) is employed to approximation of particular solution. Despite the fact that there are not nonsingular RBF general solutions available for Laplace and biharmonic problems, this study shows that the method can be successfully applied to these problems. The high-order general and fundamental solutions of Burger and Winkler equations are also first presented here.

**Introduction**: As a boundary-type RBF scheme, the method of fundamental solution (MFS), also known as regular boundary elements, attains refresh attentions in recent years [2]. Because of the use of singular fundamental solution, the MFS requires a controversial fictitious boundary outside physical domain, which effectively blocks its practical use for complex geometry problems. Chen and Tanaka [1] recently developed a boundary knot method (BKM), where the perplexing artificial boundary is eliminated via the nonsingular general solution. Just like the MFS and dual reciprocity BEM (DR-BEM) [3], the BKM also applies the DRM to approximate the particular solution. The method is symmetric, spectral convergence, integration-free, meshfree and easy to learn and implement, and successfully applied to Helmholtz, convection-diffusion, and Winkler plate problems. Unfortunately, the nonsingular RBF general solutions of Laplace and biharmonic operators are a constant rather than the RBF. Based on some physical investigations, this paper presented a few simple strategies to apply the BKM to these problems without losing its merits.

**BKM for Laplacian**: For a complete description of the BKM see ref. 2. Here we begin with a Laplace problem

$$\nabla^2 u = f(x), \quad x \in \Omega, \tag{1}$$

$$u(x) = R(x), \quad x \subset S_u, \qquad \frac{\partial u(x)}{\partial n} = N(x), \quad x \subset S_T, \tag{2a,b}$$

where $x$ means multi-dimensional independent variable, and $n$ is the unit outward normal. The governing equation (1) can be restated as

$$\nabla^2 u + \delta^2 u = f(x) + \delta u \quad \text{or} \quad \nabla^2 u - \delta^2 u = f(x) - \delta u, \tag{3a,b}$$

where $\delta$ is an artificial parameter. Eqs. (3a,b) are respectively Helmholtz and diffusion-reaction equations. Their zero and high order general solutions [3] are

$$u_m^\#(r) = Q_m (\gamma r)^{-n/2+1+m} J_{n/2-1+m}(\gamma r), \quad \text{and} \quad u_m^\#(r) = Q_m (\gamma r)^{-n/2+1+m} I_{n/2-1+m}(\gamma r), \quad n \geq 2, \tag{4}$$

where $n$ is the dimension of the problem; $Q_m = Q_{m-1}/(2*m*\gamma^2)$, $Q_0 = 1$; $m$ denotes the order of general solution; $J$ and $I$ represent the Bessel and modified Bessel function of the first kind. The solution of the problem can be split as the homogeneous and particular solutions

$$u = u_h + u_p, \tag{5}$$

The latter satisfies the governing equation but not boundary conditions. To evaluate the particular solution, the inhomogeneous term is approximated by

$$f(x) \cong \sum_{j=1}^{N+L} \beta_j \varphi(r_j), \tag{6}$$

where $\beta_j$ are the unknown coefficients. $N$ and $L$ are respectively the numbers of knots on the domain and boundary. The use of interior points is usually necessary to guarantee the accuracy and convergence of the BKM solution. $r_j = \|x - x_j\|$ represents the Euclidean distance norm, and $\varphi$ is the radial basis function.

By forcing approximation representation (5) to exactly satisfy governing equations at all nodes, we can uniquely determine

$$\beta = A_\varphi^{-1}\{f(x_i)\}, \tag{7}$$

where $A_\varphi$ is nonsingular RBF interpolation matrix. Then we have

$$u_p = \sum_{j=1}^{N+L} \beta_j \phi(\|x - x_j\|), \tag{8}$$

where the RBF $\phi$ is related to the RBF $\varphi$ through governing equations. In this study, we chose the first and second order general solutions as the RBFs $\phi$ and $\varphi$.

On the other hand, the homogeneous solution $u_h$ has to satisfy both governing equation and boundary conditions. By means of nonsingular general solution, the unsymmetric and symmetric BKM expressions are given respectively by

$$u_h(x) = \sum_{k=1}^{L} \alpha_k u_0^\#(r_k), \quad u_h(x) = \sum_{s=1}^{L_d} a_s u_0^\#(r_s) - \sum_{s=L_d+1}^{L_d+L_N} a_s \frac{\partial u_0^\#(r_s)}{\partial n}, \tag{9a,b}$$

where $k$ is the index of source points on boundary, $\alpha_k$ are the desired coefficients; $n$ is the unit outward normal as in boundary condition (2b), and $L_d$ and $L_N$ are respectively the numbers of knots on the Dirichlet and Neumann boundary surfaces. The minus sign associated with the second term is due to the fact that the Neumann condition of the first order derivative is not self-adjoint. Hereafter we only consider the symmetric BKM for the brevity. In terms of representation (9b), the collocation analogue equations (3a) (or (3b)) and (2a,b) are written as

$$\sum_{s=1}^{L_d} a_s u_0^\#(r_{is}) - \sum_{s=L_d+1}^{L_s+L_N} a_s \frac{\partial u_0^\#(r_{is})}{\partial n} = R(x_i) - u_p(x_i), \tag{10}$$

$$\sum_{s=1}^{L_d} a_s \frac{\partial u_0^\#(r_{js})}{\partial n} - \sum_{s=L_d+1}^{L_d+L_N} a_s \frac{\partial^2 u_0^\#(r_{js})}{\partial n^2} = N(x_j) - \frac{\partial u_p(x_j)}{\partial n}, \tag{11}$$

$$\sum_{s=1}^{L_d} a_s u_0^\#(r_{ls}) - \sum_{s=L_d+1}^{L_d+L_N} a_s \frac{\partial u_0^\#(r_{ls})}{\partial n} = u_l - u_p(x_l). \tag{12}$$

Note that $i$, $s$ and $j$ are reciprocal indices of Dirichlet ($S_u$) and Neumann boundary ($S_\Gamma$) nodes. $l$ indicates response knots inside domain $\Omega$. Then we can employ the obtained expansion coefficients $\alpha$ and inner knot solutions $u_l$ to calculate the BKM solution at any other knots.

If the inhomogeneous solution $u_p$ is simply ignored (i.e., let $u=u_h$) when $\delta$ is reasonably small, the above procedure for particular solution is omitted. We only need to solve analog equations

$$\sum_{s=1}^{L_d} a_s u_0^\#(r_{is}) - \sum_{s=L_d+1}^{L_s+L_N} a_s \frac{\partial u_0^\#(r_{is})}{\partial n} = R(x_i), \tag{13}$$

$$\sum_{s=1}^{L_d} a_s \frac{\partial u_0^\#(r_{js})}{\partial n} - \sum_{s=L_d+1}^{L_d+L_N} a_s \frac{\partial^2 u_0^\#(r_{js})}{\partial n^2} = N(x_j), \tag{14}$$

This strategy is called simplified BKM hereafter which can be understood that the use of nonsingular general solutions of Helmholtz-like operators with small characteristic parameter approximates the constant general solution of the Laplace and harmonic operators. I found that the simplified BKM is not stable for irregular geometry since the poor accuracy appears at very few nodes. However, it is noted that the strategy can produce very accurate solutions for regular geometry. For instance, the $L_2$ relative error norm at 495 nodes of an ellipse for the following 2D Laplace problem (15) by the simplified BKM using 9 nodes is 5.3e-3.

**Numerical results and discussions**: Figs. 1 and 2 show the tested 2D and 3D irregular geometries, where the 3D ellipsoid cavity locates at the center of the cube with the characteristic lengths 3/8, 1/8 and 1/8. Except Neumann boundary conditions on $x=0$ surface of 3D case, the otherwise boundary are all Dirichlet type. The tested 2D and 3D examples have accurate solutions

$$u = x^3 y - xy^3 + 10x + 10, \qquad u = x^3 yz - 2xy^3 z + xyz^3 + 10x + 10. \qquad (15a,b)$$

The BKM $L_2$ norms of relative errors are displayed in Table 1. Note that the BKM only uses 9 boundary knots ($\delta=0.1$) for 2D case and 66 ones ($\delta=0.2$) for 3D case. The corresponding $L_2$ norms of relative errors are calculated at 492 sample nodes for 2D and 1000 sample nodes for 3D. The absolute error is taken as the relative error if the absolute value of the solution is less than 0.001. One can find that the present BKM methodology is very simple, accurate and efficient compared with other methods, especially for complicated geometry domain.

Although we do not use inner knots in the present test, a few inner nodes are usually necessary in practical use to significantly improve the solution accuracy and stability (i.e., insensitive to artificial parameter $\delta$).

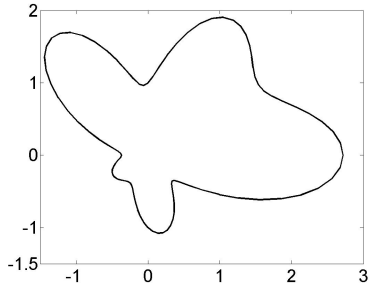 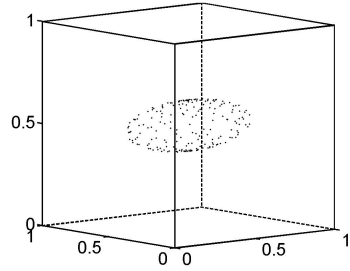

Fig. 1. A 2D irregular geometry    Fig. 2. A cube with an ellipsoid cavity

Table 1. $L_2$ norm of relative errors for 2D and 3D Laplace problems with the general solution of Helmholtz (H) and modified Helmholtz (MH) operators.

| BKM (H) | BKM (MH) | BKM (H) | BKM (MH) |
|---|---|---|---|
| 1.1e-3 (2D) | 1.5e-3 (2D) | 5.2e-3 (3D) | 1.3e-3 (3D) |

**Biharmonic problems and high order general solutions of Berger and Winkler equations**: We can use the general solutions of vibration plate, Winkler plate and Burger equation of finite deflection of plate to approximate the constant general solution of the biharmonic operator. We list these general solutions in Table 2, where ber and bei respectively represent the Kelvin and modified Kelvin functions of the first kind. Among them, it is believed that those of Winkler plate and Burger plate are first presented here.

Table 2. $M$-order general solutions of vibration plate, Winkler equation and Burger equation, where $n=2,3$ denotes dimensionality.

| | Operators | General solutions ($m=0,1,2…$) |
|---|---|---|
| Vibration plate | $L\{u\}=\nabla^4 u - \lambda^2 u$ | $u_m^\#(r) = \left(r\sqrt{\lambda}\right)^{-n/2+1+m}\left(A_m J_{n/2-1+m}\left(\sqrt{\lambda}r\right) + B_m I_{n/2-1+m}\left(\sqrt{\lambda}r\right)\right)$ |
| Winkler plate | $L\{u\}=\nabla^4 u + \kappa^2 u$ | $u_m^\#(r) = \left(r\sqrt{\kappa}\right)^{-n/2+1+m}\left(C_m ber_{n/2}(r\sqrt{\kappa}) + D_m bei_{n/2}(r\sqrt{\kappa})\right)$, $m$ odd $u_m^\#(r) = \left(r\sqrt{\kappa}\right)^{-n/2+1+m}\left(C_m ber_{n/2-1}(r\sqrt{\kappa}) + D_m bei_{n/2-1}(r\sqrt{\kappa})\right)$, $m$ even |
| Burger plate | $L\{u\}=\nabla^4 u - \mu^2 \nabla^2 u$ | $u_m^\#(r) = E_m r^{2m-2} + F_m (\mu r)^{-n/2+1+m} I_{n/2-1+m}(\mu r)$ |

$A_m$ to $F_m$ are constant coefficients which will analyzed in a subsequent paper. It is worth pointing out that the formulas given in Table 2 for the zero order general solution of Winkler operator is effective for up to five dimensions. The same relations hold with ber, bei replaced by the Kelvin functions of the second kind ker, kei, respectively, for fundamental solutions. The higher order fundamental solution of Burger equation is alternations of the first and second terms of corresponding general solution by the higher order fundamental solutions of Laplace and Helmholtz operators.

The fundamental solutions of Winkler plate and Burger plate are given respectively by [5,6]. For vibration plate, the small vibration frequency means the approximation to linear steady deflection described by a biharmonic equation. For a Winkler plate on an elastic foundation, the small elastic foundation coefficient indicates that the general solution is close to that of a plate no resting on an elastic foundation. Ref. 5 has actually applied a BEM strategy based on the fundamental solution of the Winkler plate to analyze the biharmonic system equation, where the inhomogeneous (particular) solution is simply omitted, similar to idea in the aforementioned simplified BKM. The solution accuracy is quite high for a regular rectangular plate.

The Burger plate equation is a simplified model of von Karman equations for nonlinear deflection of plate under large loading, which assumes the plate has not in-plane movement at the boundary. By taking small Burger parameter, the nonsingular general solution of Burger plate approaches that of the biharmonic equation for the linear thin plate.

Now it is obvious that the present approximate strategy is explicitly grounded on the physical connections between different partial differential equations. The numerical validations of the biharmonic equations will be presented in a subsequent paper.

**Remarks**: It should be pointed out that we could greatly simplify the above-given standard form of general solutions involving some special functions. Thus, the computing effort for them is trivial. For Helmholtz-like problems, the BKM outperforms the DR-BEM and MFS significantly in terms of accuracy, symmetricity, efficiency, stability, and mathematical simplicity. The present study shows that the method is also very efficient for Laplace and biharmonic problems by a fairly good approximation via general solutions of Helmholtz-like operators with small system parameter to that of the corresponding non-Helmholtz-like operators. A mathematical analysis of this scheme will be given later. The major drawbacks of the BKM are severe ill-condition and costly full matrix for large system problems, which is a subject presently under investigation.